\catcode`\@=11
\newif\if@fewtab\@fewtabtrue
{\count255=\time\divide\count255 by 60
\xdef\hourmin{\number\count255}
\multiply\count255 by-60\advance\count255 by\time
\xdef\hourmin{\hourmin:\ifnum\count255<10 0\fi\the\count255}}
\def\ps@draft{\let\@mkboth\@gobbletwo
    \def\@oddfoot{\hbox to 7 cm{\tiny \versionno
       \hfil}\hskip -7cm\hfil\rm\thepage \hfil {\tiny\draftdate}}
    \def\@oddhead{}
    \def\@evenhead{}\let\@evenfoot\@oddfoot}
\def\draftdate{\number\month/\number\day/\number\year\ \ \ \hourmin }

\global\def\draftcontrol{0}
\def\citen#1{\if@filesw \immediate\write \@auxout {\string\citation{#1}}\fi%
\@tempcntb\m@ne \let\@h@ld\relax \def\@citea{}%
\@for \@citeb:=#1\do {\@ifundefined {b@\@citeb}%
    {\@h@ld\@citea\@tempcntb\m@ne{\bf ?}%
    \@warning {Citation `\@citeb ' on page \thepage \space undefined}}%
    {\@tempcnta\@tempcntb \advance\@tempcnta\@ne
    \setbox\z@\hbox\bgroup\ifcat0\csname b@\@citeb \endcsname \relax
    \egroup \@tempcntb\number\csname b@\@citeb \endcsname \relax
    \else \egroup \@tempcntb\m@ne \fi \ifnum\@tempcnta=\@tempcntb
    \ifx\@h@ld\relax \edef \@h@ld{\@citea\csname b@\@citeb\endcsname}%
    \else \edef\@h@ld{\hbox{--}\penalty\@highpenalty
    \csname b@\@citeb\endcsname}\fi
    \else \@h@ld\@citea\csname b@\@citeb \endcsname \let\@h@ld\relax \fi}%
\def\@citea{,\penalty\@highpenalty\hskip.13em plus.13em minus.13em}}\@h@ld}
\def\@citex[#1]#2{\@cite{\citen{#2}}{#1}}%
\def\@cite#1#2{\leavevmode\unskip\ifnum\lastpenalty=\z@\penalty\@highpenalty\fi%
  \ [{\multiply\@highpenalty 3 #1%
  \if@tempswa,\penalty\@highpenalty\ #2\fi}]}   %
\makeatother 
\catcode`\@=12

\def\A             {Algebra}
\def\ala           {\mathfrak A}
\def\alg           {algebra}
\def\Atop          {\mbox{$A_{\rm top}$}}
\def\bc            {boundary condition}
\def\be            {\begin{equation}}
\def\bearl         {\begin{array}{l}}
\def\bearll        {\begin{array}{ll}}
\def\calc          {\mbox{$\mathcal C$}}
\def\calh          {\mbox{$\mathcal H$}}
\def\cals          {\mbox{$\mathcal S$}}
\def\cats          {categories}
\def\cft           {conformal field theory}

\def\cfts          {conformal field theories}
\def\chii          {\raisebox{.15em}{$\chi$}}
\def\cir           {\,{\circ}\,}
\def\class         {classification}
\def\Class         {Classification }
\def\cocon         {coset construction}  
\def\complex       {\mbox{$\mathbb C$}}
\def\con           {conformal }
\def\Con           {Conformal }
\def\corfu         {correlation function}
\global\def\draftcontrol{0}
\def\dim           {{\rm dim}}

\def\ee            {\end{equation}}
\def\eear          {\end{array}}
 
\def\eps           {\varepsilon}
\def\eq            {\,{=}\,}
\newcommand\erf[1] {(\ref{#1})}
\def\findim        {fini\-te-di\-men\-si\-o\-nal}
\def\ft            {field theory}
\def\fts           {field theories}
\def\Hom           {{\rm Hom}}

\def\hy            {$\mbox{-\hspace{-.66 mm}-}$}
\def\id            {\mbox{\sl id}}
\def\ide           {identification}  

\def\ii            {{\rm i}}

\def\iN            {\,{\in}\,}
\def\ksb           {\Xi}
\newcommand\labl[1]{\label{#1}\ee \ifnum\draftcontrol=1
                   \mbox{ }\\[-12 mm]\query{#1}\\[5 mm] \fi}
\def\maccor        {McKay correspondence}

\def\M             {{\dot M}}

\def\modinv        {modular invarian}

\def\one           {{\bf1}}

\def\oti           {\,{\otimes}\,}
\def\Oti           {{\otimes}}
\def\parfu         {partition function}
\def\q             {quantum }
\def\Q             {Quantum }
\def\QFT           {Quantum Field Theory}        
\def\qfts          {quantum field theories}
\def\rep           {representation}
\def\Rep           {Representation}

\def\smallone      {{\bf1}}
\def\tft           {topological field theory}
\def\tfts          {topological field theories}
\def\twodim        {two-di\-men\-si\-o\-nal}
\def\Vee           {^{\!\vee}}

\def\wzwm          {WZW model}
\def\zet           {\mbox{$\mathbb Z$}}

\documentclass[12pt]{article}
\usepackage{amssymb,amsfonts,epsf,color,colordvi}
\usepackage[dvips]{graphics} 
\begin{document}

\begin{flushright}  {~} \\[-12mm]
{\sf hep-th/0110133}\\[1mm]{\sf PAR-LPTHE 01-45}\\[1mm]{ESI-1103}\\[1mm]
{\sf October 2001} \end{flushright}

\begin{center} \vskip 14mm
{\Large\bf CONFORMAL CORRELATION FUNCTIONS,}\\[3mm]
{\Large\bf FROBENIUS ALGEBRAS AND TRIANGULATIONS}\\[15mm] {\large 
J\"urgen Fuchs$\;^1$ \ \ Ingo Runkel$\;^2$ \ \ Christoph Schweigert$\;^2$}
\\[8mm]
$^1\;$ Institutionen f\"or fysik~~~~{}\\
Universitetsgatan 1\\ S\,--\,651\,88\, Karlstad\\[5mm]
$^2\;$ LPTHE, Universit\'e Paris VI~~~{}\\
4 place Jussieu\\ F\,--\,75\,252\, Paris\, Cedex 05
\end{center}
\vskip 13mm
\begin{quote}{\bf Abstract}\\[1mm]
We formulate two-dimensional rational conformal field theory as a 
natural generalization of two-dimensional lattice topological field
theory. To this end we lift various structures from complex vector spaces
to modular tensor categories. The central ingredient is a special
Frobenius algebra object $A$ in the modular category that encodes the 
Moore\hy Seiberg data of the underlying chiral CFT. Just like for 
lattice TFTs, this algebra is itself not an observable quantity.
Rather, Morita equivalent algebras give rise to equivalent theories.
Morita equivalence also allows for a simple understanding of T-duality.
\\
We present a construction of correlators, based on a triangulation of
the world sheet, that generalizes the one in lattice TFTs. These 
correlators are modular invariant and satisfy factorization rules. 
The construction works for arbitrary orientable world sheets, in 
particular for surfaces with boundary.
Boundary conditions correspond to representations of the algebra $A$.
The partition functions on the torus and on the annulus provide
modular invariants and NIM-reps of the fusion rules, respectively.
\end{quote}
\vfill
\newpage

\section{Introduction}

Attempts to understand the spectrum of bulk fields in \twodim\ conformal
field theories gave rise to the formulation of a mathematical problem: Classify
modular invariant torus partition functions. Considerable effort has
been spent on this problem over the past 15 years. In theories of closed
strings, modular invariance of the partition function guarantees the absence
of anomalies in the low-energy effective field theory. Modular invariance is 
a much stronger property than anomaly freedom, i.e.\ string theory
imposes strictly stronger conditions on the field content than
field theory. It comes therefore as a disappointment that there exist
modular invariants that obey all the usual constraints -- positivity,
integrality, and uniqueness of the vacuum -- but are nevertheless unphysical
(see e.g.\ \cite{scya5,fusS}). Thus, albeit a mathematically
well-posed problem, classifying modular invariants is not exactly what is
desired from a physical point of view.

The study of the open string field content of \cfts\ resulted in the
formulation of a similar problem \cite{sasT2,pezu4}: Classify NIM-reps, that 
is, representations of the fusion rules by matrices with non-negative integral 
entries. Again, this classification yields (plenty of) spurious solutions that
cannot appear in a consistent \cft\ \cite{gann17}.

Motivated by these observations we pose the following questions: First, what 
is the correct structure that allows
to classify full rational conformal field theories with given Moore\hy Seiberg
data? And second, how does this structure determine the correlation functions
of the full CFT, in particular how does one obtain a modular invariant 
partition function (as the 0-point correlator on the torus) and NIM-reps
(as the 0-point correlators on the annulus)? In this note
we present a natural structure that allows to construct
correlation functions and that can be expected to arise in every RCFT.
The structure in question is the one of a {\em symmetric special Frobenius 
algebra\/} in a {\em modular tensor category\/}. The resulting correlation 
functions are modular invariant and satisfy factorization rules.

Neither the presence of this structure nor its relevance to the questions 
above is a priori obvious. Moreover, as the choice of words indicates, to 
formulate our prescription we need to employ some tools which are not 
absolutely standard. In particular we work in the context of modular tensor 
categories. Their relevance to our goal emerges from the fact that they provide 
a powerful graphical calculus and a convenient basis-free formalization of the 
Moore\hy Seiberg data of a chiral \cft, like braiding and fusing 
matrices and the modular S-matrix. A well-known example of a modular
tensor category, which can serve as a guide to the general theory, is the
category of \findim\ vector spaces over the complex numbers. This category is 
also relevant to the analysis of two-dimensional lattice topological theory 
\cite{fuhk}, so that one may suspect a relation between those theories and 
our prescription. Indeed, one of our central observations is that much of
the structure of (rational) conformal field theories can be understood
in terms of constructions familiar from lattice TFT, provided that one
adapts them so as to be valid in general modular tensor categories as well.

The rest of the paper is organized as follows.
In Section 2 we provide the necessary background on modular tensor categories 
and motivate the appearance of Frobenius algebras. In Section 3 we develop the
representation theory of these algebras. In Section 4 it is shown how to obtain
consistent torus and annulus partition functions from a symmetric special 
Frobenius algebra. In Section 5 we give a prescription for general correlators 
that generalizes the one for \twodim\ lattice TFTs in the category of vector 
spaces. It turns out that the Frobenius algebra itself is not observable; 
Morita equivalent algebras give rise to equivalent theories. When combined 
with orbifold techniques, this fact allows for a general understanding of 
T-duality in rational CFT. This is discussed in Section 6. 
Section 7 contains our conclusions.

A much more detailed account of our results, 
including proofs, will appear elsewhere.

\section{Frobenius algebras}

As already pointed out, our considerations are formulated in the language of
modular tensor categories \cite{TUra}, a formalization of Moore\hy Seiberg 
\cite{Mose} data. A modular tensor category $\calc$
may be thought of as the category of representations of some chiral
algebra $\ala$, which in turn correspond to the primary fields of a chiral 
\cft. Accordingly the data of $\calc$ can be summarized as follows.
The (simple) {\em objects\/} of $\calc$ are the (irreducible) \rep s
of $\ala$, and the {\em morphisms\/} of $\calc$ are $\ala$-intertwiners. For 
a rational conformal field theory, the category is semisimple, so that every 
object is a finite direct sum of simple objects. There is a tensor product 
$\otimes$ which corresponds to the
(fusion) tensor product of $\ala$-\rep s, and the
vacuum \rep\ (identity primary field) provides a unit element $\one$
for this tensor product. The existence of conjugate
$\ala$-representations gives rise to a {\em duality\/} on $\calc$,
in particular each object $V$ of $\calc$ has a dual object $V^\vee$.
The (exponentiated) conformal weight provides a {\em twist\/} on
$\calc$: To every object there is associated a twist endomorphism $\theta_V$; 
for a simple object $V$ corresponding to a primary field of conformal weight 
$\Delta_V$, $\theta_V$ is a multiple of the identity morphism, $\exp(-2\pi
\ii\Delta_V)\,\id_V$. Finally, the presence of braid group statistics in
two dimensions is accounted for by a {\em braiding\/}, i.e.\ for every pair 
$V,W$ of objects there is an isomorphism $c_{V,W}\iN\Hom(V\Oti W,W\Oti V)$
that `exchanges' the two objects. The modular S-matrix of the CFT is obtained
from the trace of the endomorphisms $c_{V,W}c_{W,V}$ with simple $V,W$. 

These data are subject to a number of axioms that can be
understood as formalizations of various properties of primary fields in
rational CFT (see e.g.\ appendix A of \cite{fffs3}). Essentially, the axioms
guarantee that these morphisms can be visualized via ribbons and that the
graphs obtained by their composition share the properties of the corresponding
ribbon graphs, and they also include the requirement that the S-matrix is 
invertible. (That one really must use ribbons rather than lines is closely
related to the fact that Wilson lines in Chern\hy Simons gauge theories
require a framing.)

A particularly simple example of a modular
tensor category is the category of \findim\ complex vector spaces.
It has a single isomorphism class of simple objects -- the class of the
one-dimensional vector space \complex\ -- and has trivial twist
and braiding. In \cft, this category arises for meromorphic models, i.e.\ models
with a single primary field, such as the $E_8$ WZW theory at level 1.
It is also the category in which one discusses two-dimensional lattice
topological theory, and as we will discuss below, this is not a coincidence.

\medskip

Central to our construction are certain objects in modular tensor categories
that can be endowed with much further structure. To get an idea on the role
of these objects, let us collect a few more results from CFT. First, recall that
modular invariants are either of `extension type' or of `automorphism type',
or a combination thereof \cite{mose2}. 
A simple example of extension type arises for the $sl(2)$ WZW theory at level 4:
  \be Z(\tau) = |\chii_0(\tau) + \chii_4(\tau)|^2 + 2\, |\chii_2(\tau)|^2 \,.
  \ee
This invariant corresponds to the conformal embedding of $sl(2)$ level 4 into 
$sl(3)$ at level 1, and can therefore be interpreted as follows: The vacuum 
sector of the extended theory $sl(3)_1$ gives rise to a reducible sector
$A\eq(0)\,{\oplus}\,(4)$ of the $sl(2)_4$ theory. In category theoretic terms, 
$A$ is a reducible object of the modular tensor category $\calc\,{\equiv}\,
\calc(sl(2)_4)$. Now $A$ is not merely an object, but comes with plenty of 
additional structure. First, the operator product provides an associative 
multiplication on $A$. Second, since in a chiral algebra the vacuum is unique,
the $sl(2)_4$-descendants of the vacuum provide a distinguished subobject 
$(0)$ of $A$.  Finally, crossing symmetry of the 4-point conformal blocks of 
the vacuum on the sphere, which relates the s-channel and the t-channel, 
results in additional properties of the product on $A$.

The formalization of these properties of $A$ as an object of $\calc$
reads as follows \cite{fuSc16}. There is a multiplication morphism 
$m\iN\Hom(A\Oti A,A)$ that is associative and for which there exists a unit 
$\eta\iN\Hom(\one,A)$. There exists a coassociative coproduct $\Delta\iN
\Hom(A,A\Oti A)$ as well, along with a co-unit $\eps\iN\Hom(A,\one)$.
It is convenient to depict these morphisms as follows:
  \be  \begin{picture}(333,45)(0,0)
  \put(0,16)  {$m\;=$}
  \put(34,0)  {\begin{picture}(0,0)(0,0)
              \scalebox{.55}{\includegraphics{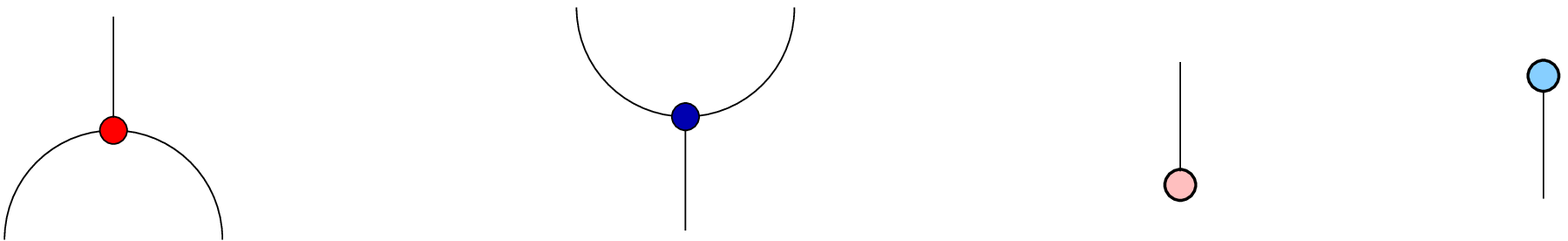}} \end{picture}}
  \put(106,16){$\Delta\;=$}
  \put(212,16){$\eta\;=$}
  \put(281,16){$\eps\;=$}
  \end{picture} \labl{def--mdee}
Such pictures are to be read from bottom to top, and composition of morphisms
amounts to concatenation of lines; a single line labelled by an object $V$
stands for the identity morphism $\id_V\iN{\rm End}(V)\,{\equiv}\,\Hom(V,V)$.
Lines labelled by the tensor unit $\one$ can be omitted since ${\rm End}(\one)
\eq\complex$ so that $\id_\smallone\eq1$. In \erf{def--mdee} we have suppressed
the label $A$ which decorates each of the lines of the picture. (Also, throughout
the paper the fattening of the lines to ribbons is implicitly understood.)

The (co-)associativity and (co-)unit properties, i.e.
  \be \bearll
  m\circ (m\oti\id_A) = m \circ (\id_A\oti m) \,, &
  m \circ(\eta\oti \id_A)  = \id_A = m \circ (\id_A\oti \eta) \,,
  \\{}\\[-.6em]
  (\Delta\oti\id_A)\circ\Delta = (\id_A\oti\Delta)\circ\Delta \,, \quad\ &
  (\eps\oti\id_A) \circ\Delta = \id_A = (\id_A\oti\eps) \circ \Delta \,,
  \eear \ee
are then drawn as
  \be  \begin{picture}(300,106)(0,0)
  \put(0,0)   {\begin{picture}(0,0)(0,0)
              \scalebox{.38}{\includegraphics{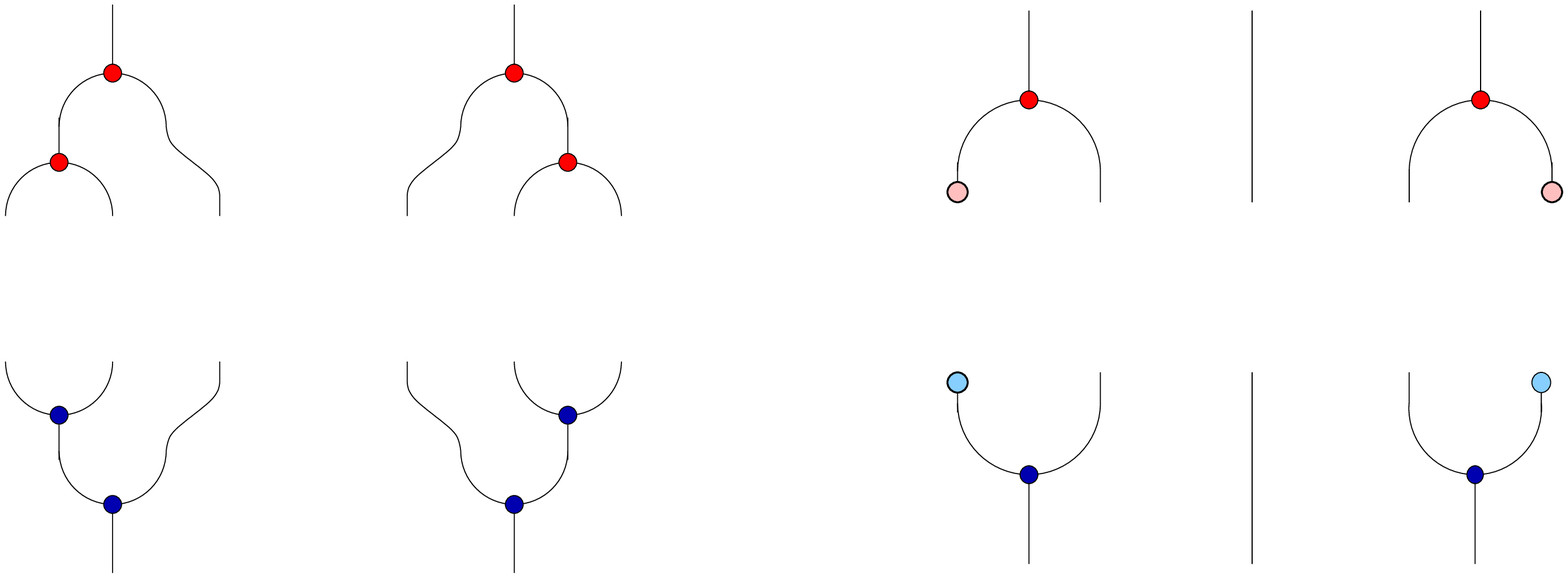}} \end{picture}}
  \put(56,19) {$=$}
  \put(56,85) {$=$}
  \put(221,19){$=$}
  \put(221,85){$=$}
  \put(252,19){$=$}
  \put(252,85){$=$}
  \end{picture} \labl{ass-coass}
Further, the crossing symmetry is taken into account by demanding $A$ to be
a {\em Frobenius algebra\/}, i.e. to satisfy
  \be  \begin{picture}(230,85)(0,0)
  \put(0,0)   {\begin{picture}(0,0)(0,0)
              \scalebox{.38}{\includegraphics{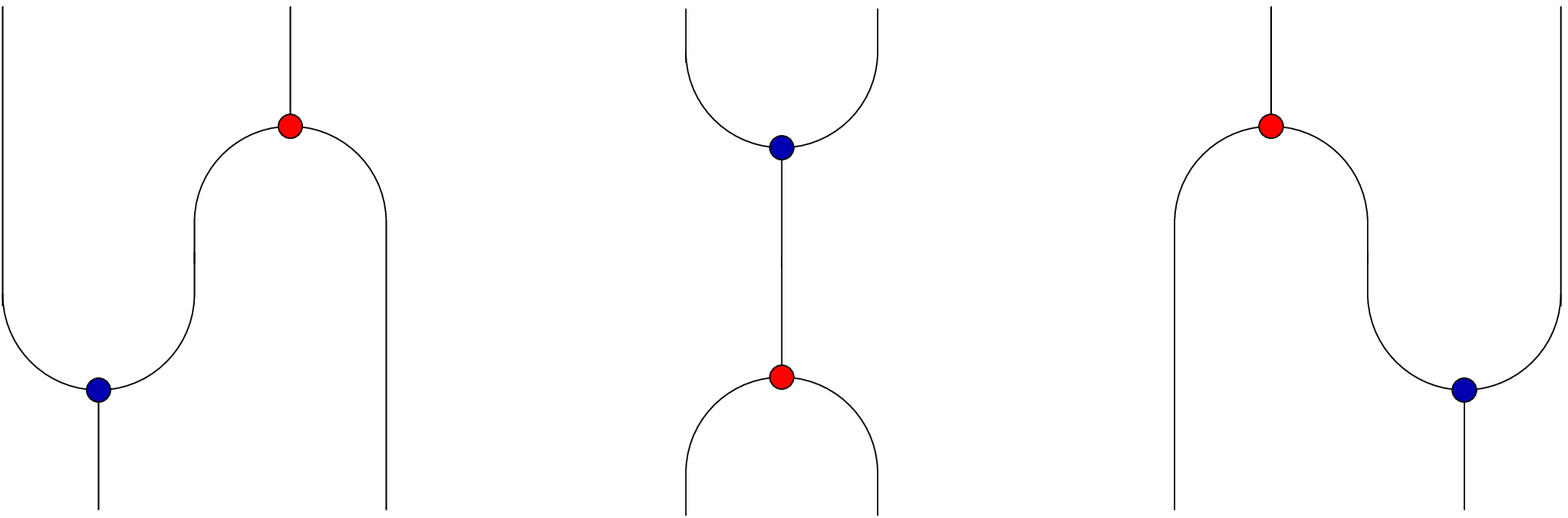}} \end{picture}}  
  \put(74,33) {$=$}
  \put(140,33){$=$}
  \end{picture} \labl{frobenius}
We also require $A$ to be a {\em special\/} Frobenius algebra, i.e.\ impose 
that (after suitably fixing the normalization of the co-unit)
  \be \eps\cir \eta = \dim(A)\, \id_\smallone \qquad\mbox{and}\qquad
  m\cir\Delta = \id_A \,,  \labl5
where $\dim(V)$ stands for the quantum dimension of an object $V$, which is
defined as the trace of the identity morphism $\id_V$.
To account for the uniqueness of the vacuum, one would impose that
$\dim\,\Hom(\one,A)\eq1$, a property that has been termed {\em haploidity\/}
of $A$ in \cite{fuSc16}. But with an eye on the relation with \twodim\
lattice TFT we prefer to relax this requirement and only require that $A$ 
is {\em symmetric\/} in the sense that the morphisms
  \be  \begin{picture}(250,80)(0,0)
  \put(47,0)  {\begin{picture}(0,0)(0,0)
              \scalebox{.38}{\includegraphics{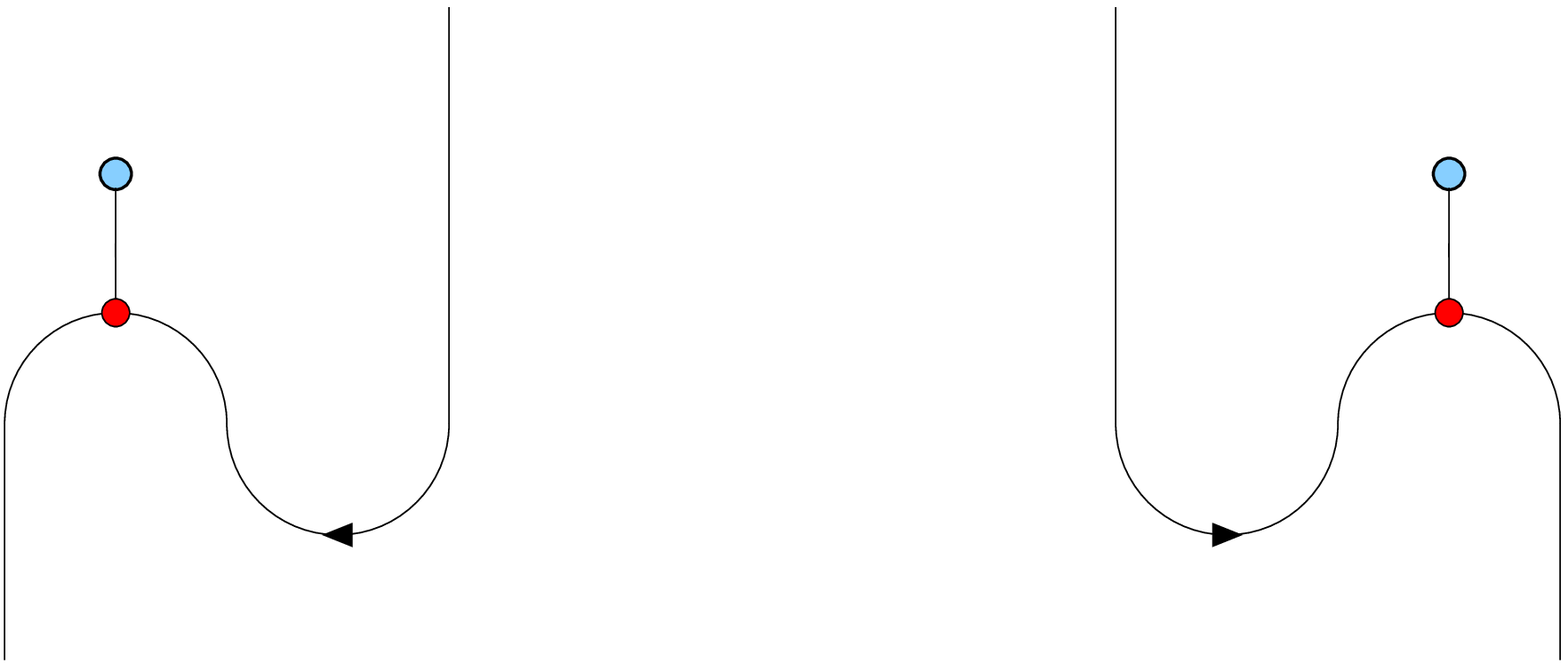}} \end{picture}}
  \put(0,31)  {$\Phi_1\;:=$}
  \put(39.6,-1) {\scriptsize$A$}
  \put(104.6,77){\scriptsize$A\Vee$}
  \put(135,31)  {$\Phi_2\;:=$}
  \put(173,77)  {\scriptsize$A\Vee$}
  \put(240.9,-1){\scriptsize$A$}
  \end{picture} \labl{Phi12}
from $A$ to its dual $A^\vee$ are isomorphisms and coincide. 
It can be shown that haploidity implies symmetry, hence this is indeed a 
weaker requirement. But as it turns out, it still leads to a meaningful theory.

The presence of a braiding allows us to introduce a notion of commutativity: 
An algebra $A$ is called commutative iff $m\cir c_{A,A}\eq m$. But it is 
well-known that certain off-diagonal modular invariants like the $D_{\rm odd}$
series of $sl(2)$ correspond to a hidden chiral superalgebra, which is not 
commutative. Correspondingly we do {\em not\/} impose commutativity of $A$.

\medskip

Let us now present more examples. The category of vector spaces contains just
a single haploid algebra, the ground field $\complex$ itself. \complex\ is
obviously a special Frobenius algebra. Thus in a sense haploid Frobenius
algebras generalize the ground field. Now as already pointed out, instead
of haploidity we impose the weaker condition of symmetry, i.e.\ equality of
the two morphisms \erf{Phi12}. This proves to be most reasonable, as 
symmetric special Frobenius algebras in the
category of complex vector spaces are known to describe \twodim\ lattice
TFTs \cite{fuhk,duJo2,abra}. Since a topological field theory is
in particular a (rather degenerate) conformal field theory, it is gratifying
that our formalism covers this case. More generally, an algebra that
is symmetric but not haploid should be thought of as containing several
vacua that constitute a topological subsector of the theory. In fact,
this topological sector gives itself rise to the structure of a
symmetric special Frobenius algebra \Atop\ over the complex numbers.
$A$ is symmetric in the sense introduced above if and only if
\Atop\ is a symmetric Frobenius algebra in the usual \cite{CUre} sense.

For general modular tensor categories the situation is much more involved.
But one haploid special Frobenius algebra is always present, namely the
tensor unit $\one$. The associated torus \parfu\ is the charge conjugation
invariant, and the boundary conditions are precisely those which preserve
the full chiral symmetry (`Cardy case'). The construction of \corfu s
presented below then reduces to the one already given in \cite{fffs3}.

Another large class of examples is supplied by simple currents, i.e.\
simple objects $J$ of $\calc$ of quantum dimension $\dim(J)\eq1$. Simple 
currents (more precisely, their isomorphism classes) form a subgroup of the 
fusion ring of the CFT that organizes the set of primary fields into orbits.
Haploid special Frobenius algebras all of whose simple subobjects are simple
currents can be classified. As objects, they are direct sums of the form
  \be  A = \bigoplus_{J\in\mathcal H} J \,,  \labl{sc}
with $\calh$ a subgroup of the group of simple currents. But not all
objects of the form \erf{sc} can be endowed with an associative product;
such a product exists if and only if the corresponding 6j-symbols yield the
trivial class in the cohomology $H^3(\calh,\complex^\times_{})$. It follows
from proposition 7.5.4 
in \cite{FRke} that this requirement restricts $\calh$ to be a subgroup of the 
so-called `effective center', i.e.\ to consist of simple currents for which the
product of the conformal weight $\Delta_J$ and the order $N_J$ (the smallest
natural number such that $J^{\otimes N}\,{\cong}\,\one$), is an integer.

Allowed objects of the form \erf{sc} can admit several different products
$m\iN\Hom(A\Oti A,A)$; the possible products are classified
by the second cohomology $H^2(\calh,\complex^\times_{})$. It is a fortunate
fact about abelian groups that classes in $H^2(\calh,\complex^\times_{})$
are in one-to-one correspondence with alternating bihomomorphisms on
$\calh$, i.e.\ with maps $\phi$ from $\calh\,{\times\calh}\,$ to
$\complex^\times_{}$ that are homomorphisms (i.e.\ compatible with the product
on $\calh$) in each argument and that obey $\phi(J,J)\eq1$ for all $J\iN\calh$.
Employing this result, the isomorphism class of the algebra structure
$m$ can be encoded in the {\em Kreuzer\hy Schellekens bihomomorphism\/} 
(KSB) $\ksb$ \cite{krSc}, which graphically is represented as
  \be  \begin{picture}(190,106)(0,0)
  \put(17,51)    {$\ksb(J,K)$}
  \put(76,51)    {$=$}
  \put(67.7,-1)  {\scriptsize$J$}
  \put(65.5,0)   {\begin{picture}(0,0)(0,0)
                 \scalebox{.38}{\includegraphics{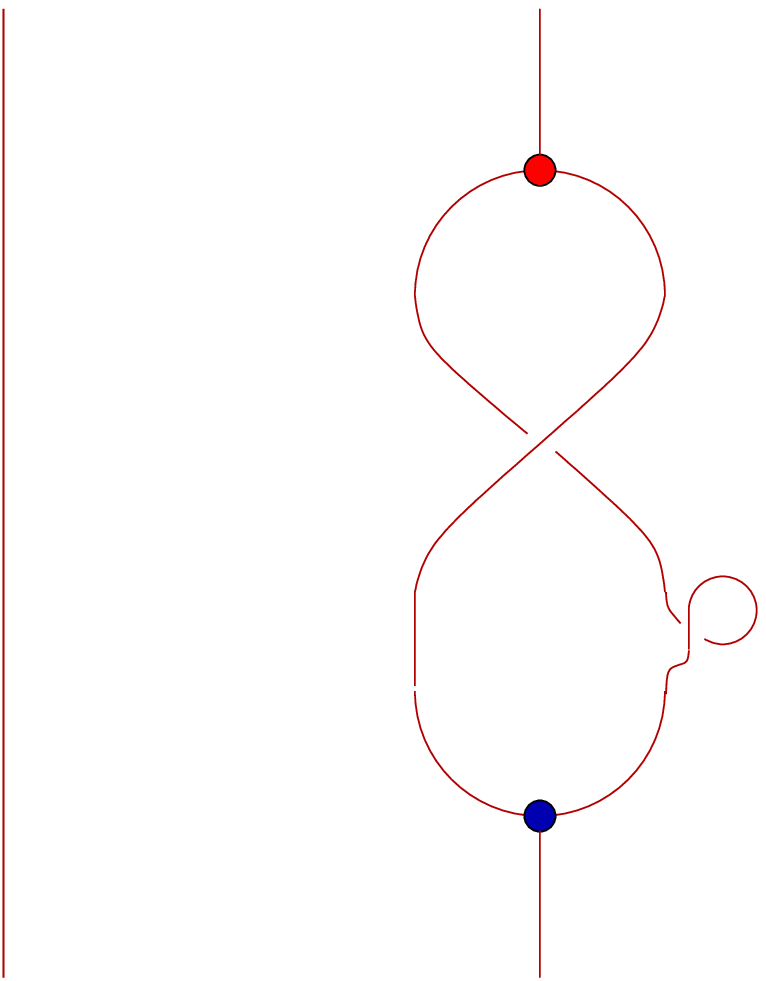}} \end{picture}}
  \put(84,25)    {\scriptsize$J{\otimes}K^\vee$}
  \put(101,73)   {\scriptsize$K$}
  \put(126.7,-1) {\scriptsize$J$}
  \put(126.7,103){\scriptsize$J$}
  \put(131.2,58) {\scriptsize$c_{J\otimes K^\vee\!\!,K}$}
  \put(150.5,38) {\scriptsize$\theta_{\!K}$}
  \end{picture} \labl{ksb}
(Here the twist morphism $\theta_K$ appears; were we drawing ribbons instead 
of lines, this would amount to a full $2\pi$ rotation of the $K$-ribbon.) The 
possible KSBs are in one-to-one correspondence with the associative products
$m$ on $\calh$; different KSBs for a simple current group $\calh$ are related 
by alternating bihomomorphisms.
The choice of a KSB has been called a choice of `discrete torsion'
in \cite{krSc}; there are indeed models where the KSB precisely describes
discrete torsion. In the case of a free boson compactified on a lattice, the
choice of a multiplication on the algebra object that characterizes the 
lattice corresponds to the choice of a background value of the $B$-field.

We finally quote examples of algebra objects for the exceptional modular 
invariants of the $sl(2)$ WZW theory. For the $E_6$-type modular invariant at
level 10, one has $A\eq(0)\,{\oplus}\,(6)$, for $E_7$ at level $16$, one finds
$A\eq(0)\,{\oplus}\,(8)\,{\oplus}\,(16)$, and for the $E_8$-type invariant at 
level $28$, $A\eq(0)\,{\oplus}\,(10)\,{\oplus}\,(18)\,{\oplus}\,(28)$
\cite{kios}.

\section{Representations}

Just like for ordinary algebras, the next step to be taken in the analysis of 
the algebra $A$ is the study of its representation theory.
Precisely as in the case of vector spaces, an $A$-representation $M$ consists
of two data: An object $\M$ of $\,\calc$, corresponding to the vector space
that underlies the module on which $A$ acts, and a representation morphism
$\rho_M\iN\Hom(A\Oti\M,\M)$ that specifies the action of $A$ on the module.
And further, these data are subject to exactly the
same constraint as in the vector space case; pictorially:
  \be  \begin{picture}(166,73)(0,0)
  \put(0,0)    {\begin{picture}(0,0)(0,0)
               \scalebox{.38}{\includegraphics{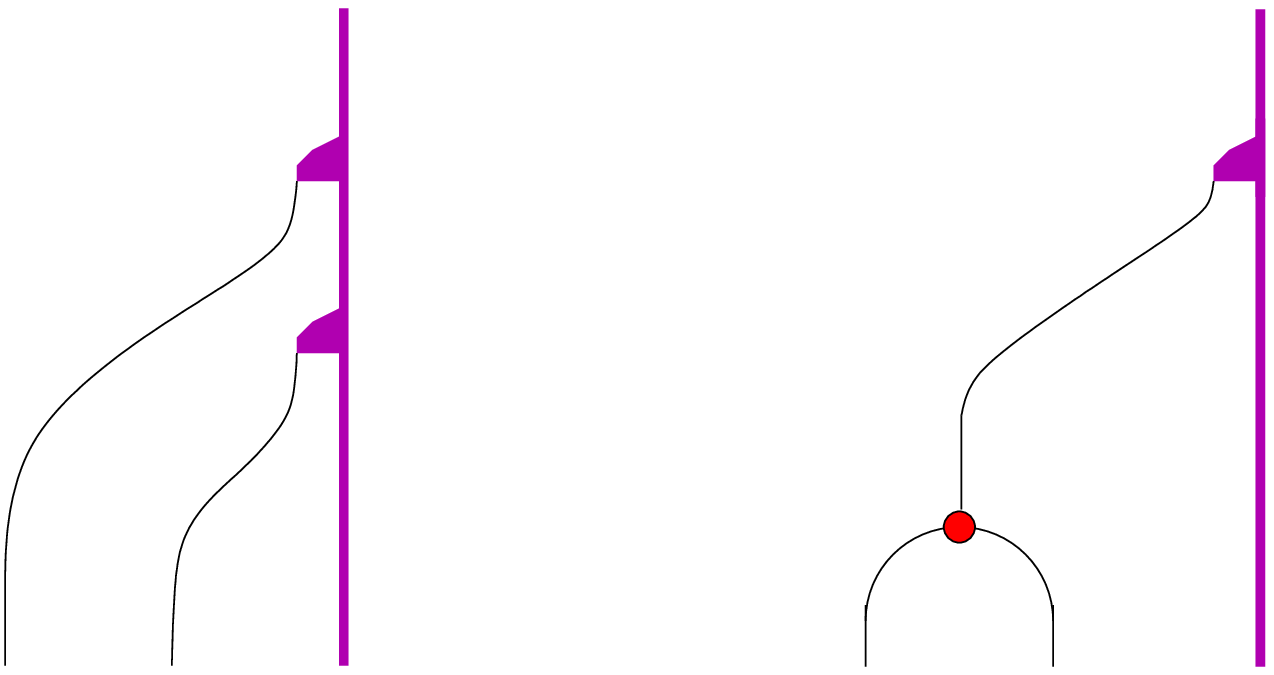}} \end{picture}}
  \put(-7.8,-1){\scriptsize$A$}
  \put(19.2,-1){\scriptsize$A$}
  \put(39.7,-1){\scriptsize$\M$}
  \put(41,37)  {\scriptsize$\rho_M$}
  \put(41,55)  {\scriptsize$\rho_M$}
  \put(73,29)  {$=$}
  \put(87.4,-1){\scriptsize$A$}
  \put(116.7,-1){\scriptsize$A$}
  \put(140,-1) {\scriptsize$\M$}
  \put(141,55) {\scriptsize$\rho_M$}
  \end{picture} \labl{rep-prop}
(In more detailed terminology, what we get this way is a {\em left\/} 
$A$-module. Right modules, as well as comodules, can be introduced analogously.
One can develop the whole theory equivalently also in terms of those.)

The representations of $A$ are in one-to-one correspondence
with conformally invariant boundary conditions. For any algebra $A$ in a 
modular tensor category, all $A$-modules are fully reducible. Irreducible
modules correspond to elementary boundary conditions, while direct sums of 
irreducible modules describe boundary conditions with non-trivial Chan\hy 
Paton multiplicities. Commutative algebras $A$ with trivial twist, i.e.\ 
$\theta_{\!A}\eq\id_A$, give rise to modular invariants of extension type. 
Irreducible modules then
fall into two different classes, {\em local\/} and {\em solitonic\/} modules.
Local modules can be characterized \cite{kios} by the fact that their twist 
is a multiple of the identity, i.e.\ that their irreducible subobjects all 
have the same conformal weight modulo integers. They correspond to boundary 
conditions that preserve all symmetries of the extension; solitonic 
representations describe symmetry breaking boundary conditions.

The standard representation theoretic tools, like induced modules and
reciprocity theorems, generalize to the category theoretic setting (see e.g.\
\cite{kios,kirI14,fuSc16}) and allow to work out the representation theory 
in concrete examples. The case of the $E_6$ modular invariant of the 
$sl(2)$ WZW theory has been presented in \cite{kios,fuSc16}; here we briefly 
comment on the representations of simple current algebras \erf{sc}. One 
starts from the observation that every irreducible module is a submodule of 
some induced module $A\Oti V$. $A$ acts on the induced module $A\Oti V$ 
from the left by multiplication, i.e.\ $\rho_{\!A\otimes V}\eq m\oti\id_V$.
We only need to consider irreducible $V$, and abbreviate the
irreducible $V_i$, with $i$ labelling the primary fields, by $i$.
As an object in $\calc$, the induced module $A\Oti i$ is just the $\calh$-orbit
$[i]$ of $i$ under fusion, with the order of the stabilizer
$\cals_i\eq\{J\iN\calh\,|\,J\Oti i\,{\cong}\,i\}$ of $i$ as the multiplicity:
  \be  A\oti i = |\cals_i|\, \bigoplus_{J\in\mathcal H/\mathcal S_i}J\Oti i \,.
  \labl{aov}
To study the decomposition of \erf{aov} into irreducible
modules, one needs to classify projectors in the endomorphism space
${\rm End}_A(A\oti i)$. This vector space possesses the structure of a
twisted group algebra over the abelian group $\cals_i$.
The twist can be computed explicitly in terms of the KSB \erf{ksb} and 
of certain gauge invariant 6j-symbols. This way one recovers the list of
boundary conditions proposed in \cite{fhssw}, which was used in \cite{fkllsw}
to compute the correct B-type boundary states in Gepner models.

\section{Partition functions}

We now demonstrate how to extract partition functions, i.e.\ torus and annulus
amplitudes, from a given symmetric special Frobenius algebra. To this end
we make use the fact that to every modular tensor category one
can associate a three-dimensional topological field theory \cite{Mose,TUra}.
Such a 3-d TFT can be thought of as a machine that associates vector spaces
-- the spaces of conformal blocks -- to two-manifolds, and linear maps 
between these vector spaces to three-manifolds. When a path integral 
formulation is available, such as the Chern\hy Simons theory in the case of 
WZW models \cite{witt27}, the vector spaces can be thought of as the
spaces of possible initial conditions, while the linear maps describe
transition amplitudes between given initial and final conditions.

More precisely, to each oriented two-manifold $\hat X$ with a finite number of
embedded arcs, which are labelled by primary fields (and certain additional
data that will not concern us here, see e.g.\ \cite{fffs3} for details), the 
TFT associates a vector space
$\calh(\hat X)$ of conformal blocks. This assignment behaves multiplicatively
under disjoint union, i.e.\ $\calh(X_1\sqcup X_2)\eq\calh(X_1)\oti\calh(X_2)$,
and assigns $\complex$ to the empty set, $\calh(\emptyset)\eq\complex$.
To an oriented three-manifold $M$ containing a Wilson graph,
with a decomposition $\partial M\eq
\partial_+M\,{\sqcup}\,\partial_-M$ of its boundary into two disjoint 
components, the TFT associates a linear map
  \be  Z(M,\partial_-M,\partial_+M):\quad \calh(\partial_-M) \to
  \calh(\partial_+M) \,.  \ee
These linear maps are multiplicative, compatible with the gluing of surfaces,
and obey a few further functoriality and naturality axioms
(see e.g.\ chapter 4 of \cite{BAki}).

As explained in \cite{fuSc6}, correlation functions of a full conformal field
theory on an arbitrary world sheet $X$ are nothing but special elements in the
space $\calh(\hat X)$ of conformal blocks on the complex cover $\hat X$ of $X$.
This is an oriented two-manifold doubly covering $X$, endowed with an
anticonformal involution $\sigma$ such that $\hat X/\sigma\eq X$. Following 
the strategy of \cite{fffs2,fffs3}, we determine the correlators using the 
`connecting three-manifold' $M_X$, which has the complex cover as its boundary,
  \be  \partial M_X = \hat X \,.  \ee
The connecting manifold is obtained as the quotient of $\hat X\,{\times}\,
[-1,1]$ by the involution that acts as $\sigma$ on $\hat X$ and as
$t\,{\mapsto}\,{-}t$ on the interval $[-1,1]$. We can and will identify the
image of $\hat X\,{\times}\,\{0\}$ in $M_X$ with $X$. The world sheet $X$
is a retract of the connecting manifold $M_X$, and hence we may think of $M_X$
as a fattening of $X$. Upon choosing an appropriate Wilson graph in $M_X$, 
the three-dimensional TFT then provides us with a map
  \be  C(X) := Z(M_X,\emptyset,\hat X):\quad \complex \to \calh(\hat X) \,,
  \ee
and it is this map that yields the correlators on $X$.

Let us now have a look at a specific situation, and explain how to obtain a 
modular invariant torus partition function from
a symmetric special Frobenius algebra. The connecting manifold for a torus
$T$ is just a cylinder over $T$, i.e.\ $M_T\eq T\,{\times}\,[-1,1]$. In this
three-manifold, we consider the following Wilson graph, labelled by $A$:
  \be  \begin{picture}(190,145)(0,0)
  \put(0,72)  {$Z\ =$}
  \put(50,0)  {\begin{picture}(0,0)(0,0)
              \scalebox{.38}{\includegraphics{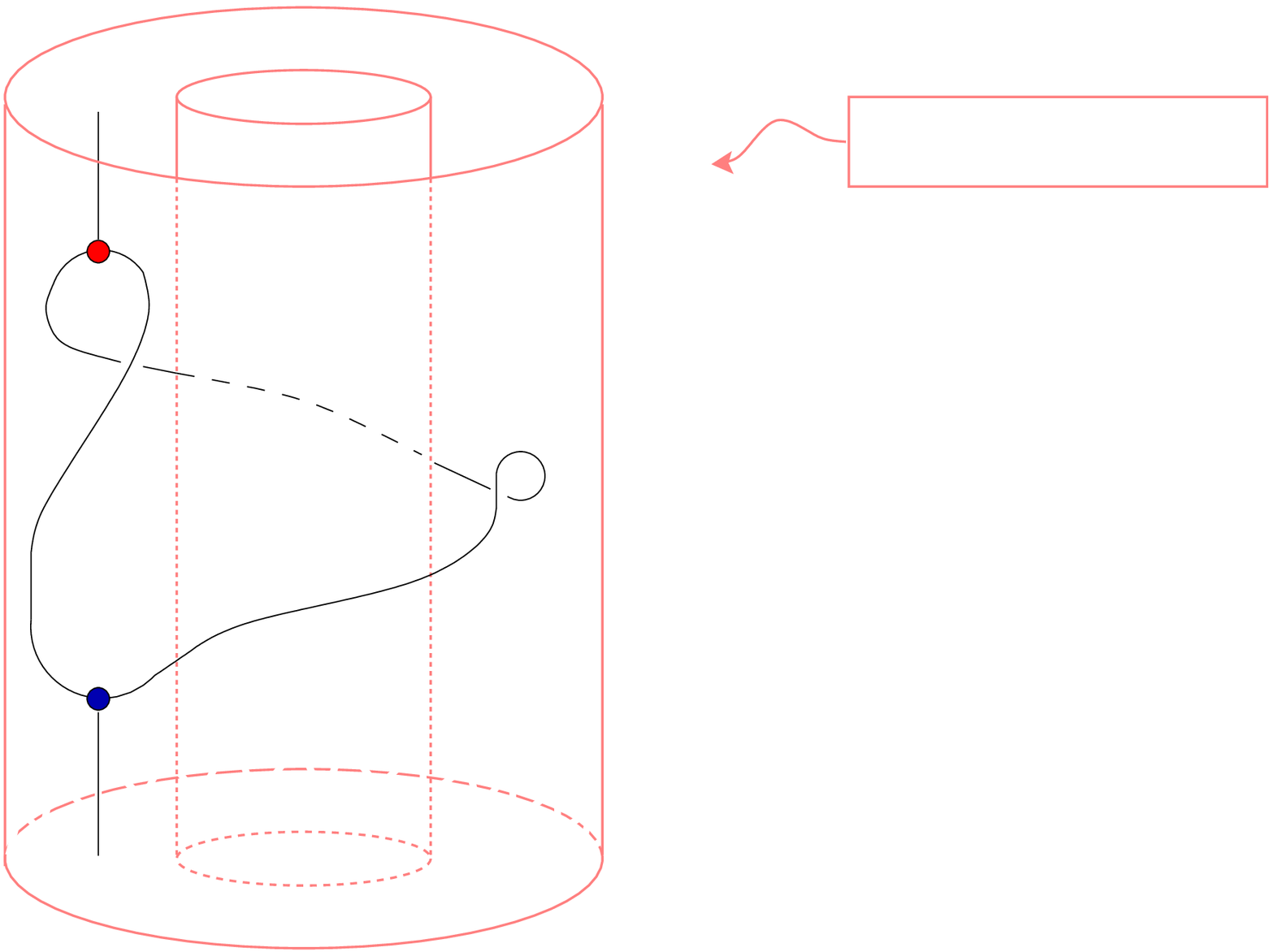}} \end{picture}}
  \put(48,87) {\scriptsize$c_{\!A,A}^{}$}
  \put(129,62){\scriptsize$\theta_{\!A}$}
  \put(183,122){\scriptsize$S^1{\times}\,[-1{,}1]\,{\times}\,S^1$}
  \end{picture} \labl{ZT}
Here (as well as in the figures \erf{Zij}, \erf{Amn} and \erf{Amni} below)
the top and bottom are to be identified, and 
the lower trivalent vertex denotes the coproduct and the upper vertex 
the product of $A$. To determine the coefficients $Z_{ij}$ of the torus \parfu,
we glue to both sides of \erf{ZT} two solid tori, containing a Wilson
line labelled by $i$ and $j$, respectively. This yields the following
Wilson graph in the three-manifold $S^2\,{\times}\,S^1$:
  \be  \begin{picture}(190,147)(0,0)
  \put(0,72)  {$Z_{ij}\ =$}
  \put(50,0)  {\begin{picture}(0,0)(0,0)
              \scalebox{.38}{\includegraphics{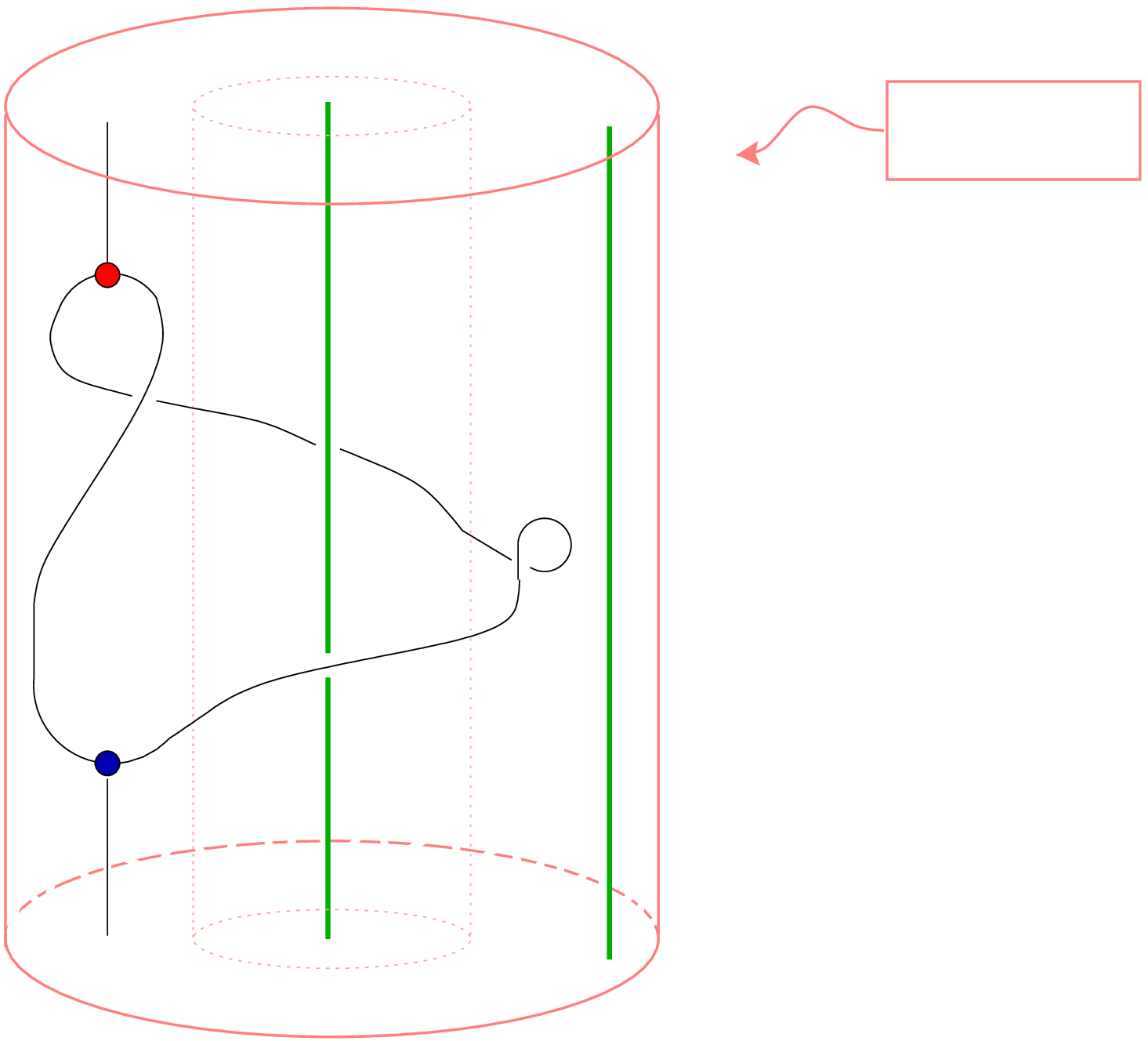}} \end{picture}}
  \put(57,13) {\scriptsize$A$}
  \put(98,13) {\scriptsize$i$}
  \put(129,13){\scriptsize$j$}
  \put(178,124.6){\scriptsize$S^2\,{\times}\,S^1$}
  \end{picture} \labl{Zij}

It is useful to interpret this result as $Z_{ij}\eq{\rm tr}\,P_{ij}$,
where $P_{ij}$ is the linear map given by the same graph in
$S^2\,{\times}\,[0,1]$. One can show that $P_{ij}$ is a projector, which
implies that the numbers $Z_{ij}$ are non-negative integers, as befits
a partition function. Furthermore, $Z_{00}$ equals the dimension of the 
center of the \complex-algebra $\Hom(\one,A)$; 
in particular, when $A$ is haploid, then the vacuum sector appears precisely 
once, $Z_{00}\eq1$. Finally, one can also prove modular invariance. We 
illustrate invariance under the U-transformation $\tau\,{\mapsto}\,\frac\tau
{\tau+1}$ (the pictures correspond to figure \erf{ZT} with the interval
$[-1,1]$ suppressed):
  \be  \begin{picture}(410,46)(0,0)
  \put(0,0)   {\begin{picture}(0,0)(0,0)
              \scalebox{.38}{\includegraphics{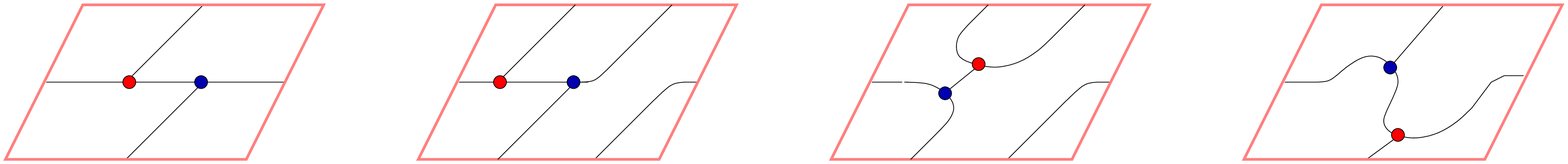}} \end{picture}}
  \put(91.3,18){\large$\stackrel U\mapsto$}
  \put(203,18){$=$}
  \put(313,18){$=$}
  \end{picture} \labl{U-trafo}
Here in the first equality the Frobenius property \erf{frobenius} of $A$ is 
used, while in the second step the periodicity of the graph is taken into
account; equality with the original picture then follows by applying the
Frobenius property once again. In the case of simple current algebras \erf{sc},
our result \erf{Zij} reproduces the formula of \cite{krSc} that expresses 
the most general simple current modular invariant in terms of the KSB and 
monodromy charges. In agreement with \cite{kios}, for commutative algebras 
$A$ with  $\theta_{\!A}\eq\id_A$ the partition function can be
shown to be of pure extension type in which only local modules appear.

For the annulus, a similar reasoning applies. The connecting manifold
looks as follows:
  \be  \begin{picture}(190,147)(0,0)
  \put(0,72)  {${\rm A}_M^{\ N}\ = $}
  \put(50,0)  {\begin{picture}(0,0)(0,0)
              \scalebox{.38}{\includegraphics{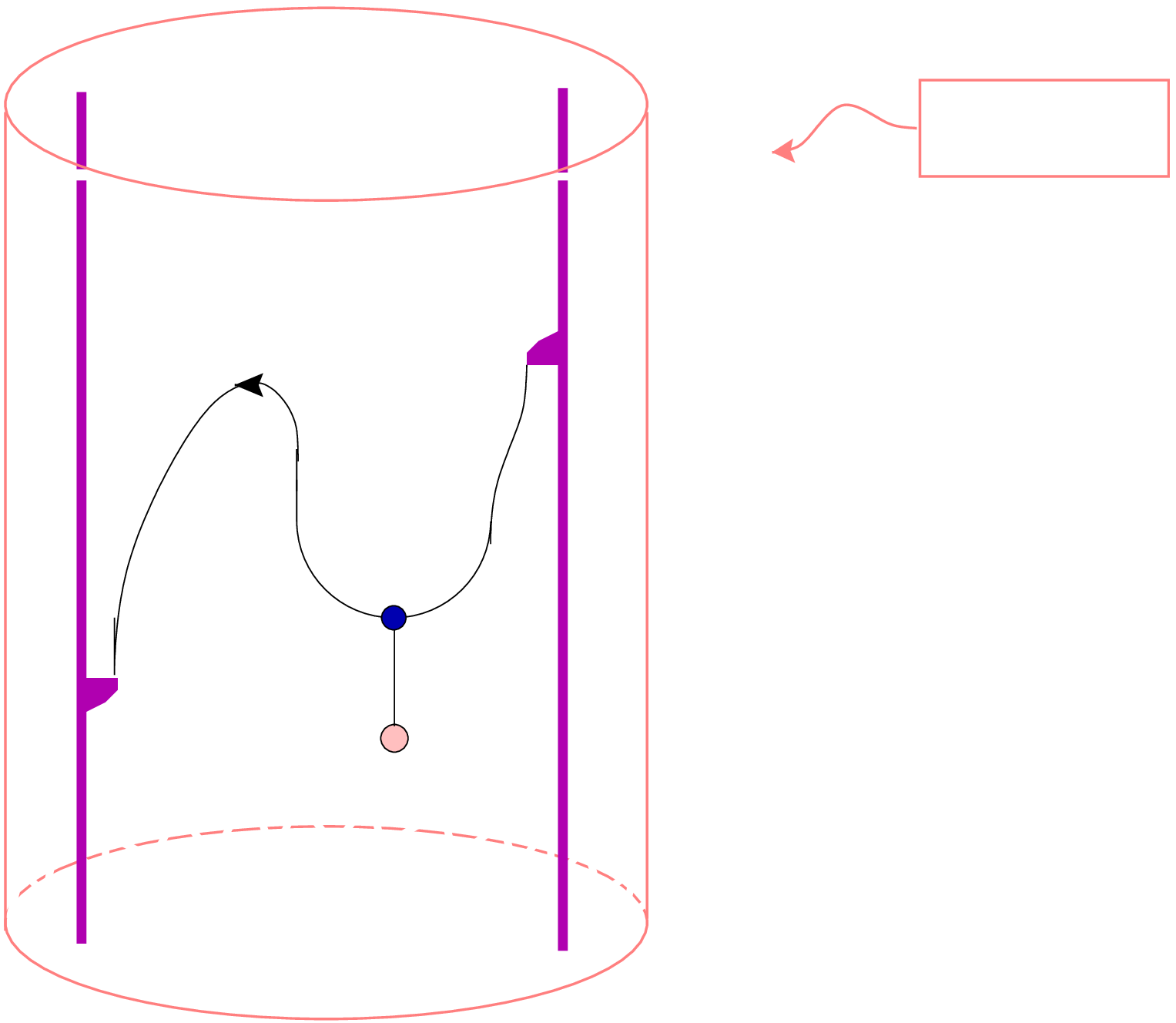}} \end{picture}}
  \put(63,11) {\scriptsize$M^\vee$}
  \put(108,51){\scriptsize$A$}
  \put(120.5,11){\scriptsize$N$}
  \put(184,125){\scriptsize$D^2\,{\times}\,S^1$}
  \end{picture} \labl{Amn}
It is a solid torus with two Wilson lines along its non-contractible cycle.
These Wilson lines have opposite orientation; each of them is is labelled by a
(left) $A$-module that characterizes the boundary condition. In addition there
are Wilson lines labelled by $A$, which are attached to the modules by
trivalent vertices that are given by the respective \rep\ morphisms. Again,
the annulus coefficients ${\rm A}_{iM}^{\ \ N}$, i.e.\ the coefficients in an
expansion of \erf{Amn} in the characters $\chii_i$, are obtained by gluing, 
in this case with a single torus containing a Wilson line labelled by $i$. 
This leads to
  \be  \begin{picture}(190,147)(0,0)
  \put(0,72)  {${\rm A}_{iM}^{\ \ N}\;= $}
  \put(50,0)  {\begin{picture}(0,0)(0,0)
              \scalebox{.38}{\includegraphics{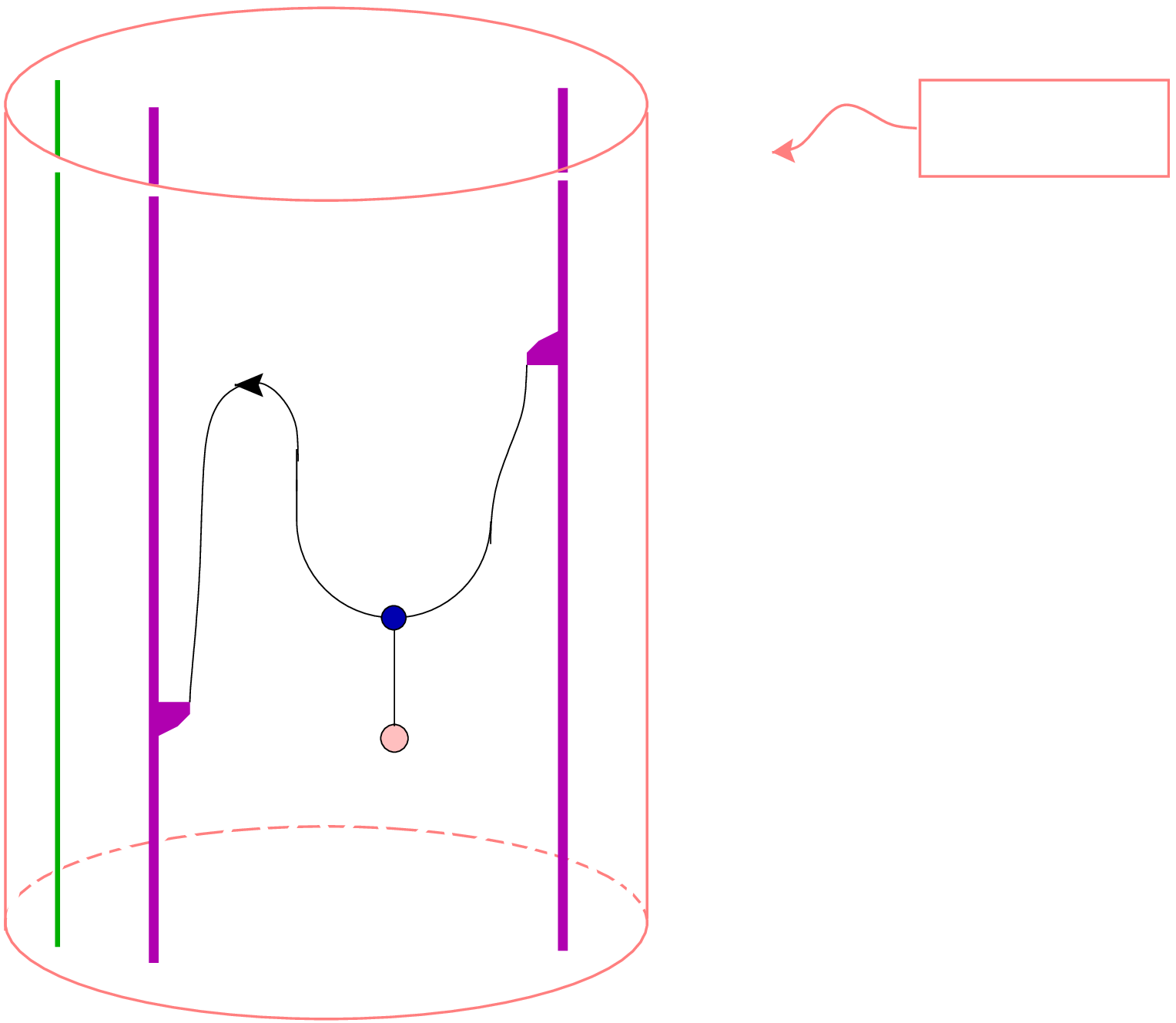}} \end{picture}}
  \put(60.8,10) {\scriptsize$i$}
  \put(73.8,9){\scriptsize$M^\vee$}
  \put(108,51){\scriptsize$A$}
  \put(120.5,11){\scriptsize$N$}
  \put(185,124.5){\scriptsize$S^2\,{\times}\,S^1$}
  \end{picture} \labl{Amni}
As in the torus case, one then shows that the annulus coefficients are 
non-negative integers. Moreover, we can show that the matrices ${\rm A}_i$
with entries $({\rm A}_i)_{\!M}^{\;N}\,{ :=}\,A_{iM}^{\ \ N}$ indeed furnish 
a NIM-rep of the fusion rules of \calc. It also follows that if the algebra 
$A$ is haploid and $M\eq N$ is an irreducible $A$-module, i.e.\ an elementary
boundary condition, then the vacuum $\one$ appears just once in the annulus,
i.e.\ ${\rm A}_{\one\,M}^{\ \ M}\eq 1$. {}From the annuli, one can read off 
the boundary states, and show that their coefficients provide
the `classifying algebra' \cite{fuSc5}. Performing a modular transformation, 
one obtains the annuli in the closed string channel; one can check that 
only fields $i$ appear that are compatible with the torus partition function
\erf{ZT}, and that they appear with the correct multiplicity $Z_{i,i^\vee}$.

\section{General amplitudes}

The results presented above can be summarized by the statement that
every symmetric special Frobenius algebra gives rise to a modular invariant
torus partition function and to a NIM-rep of the fusion rules that are
compatible with each other. To show that we even obtain a complete consistent
conformal field theory, we should construct all correlators on world sheets $X$
of arbitrary genus, including an arbitrary number of boundary components,
and show that they are invariant under the relevant mapping class group and
possess the correct factorization properties. (In the present note, we
restrict ourselves to orientable world sheets. The unorientable
case requires additional structure, so as to account for the possibility
of having several different Klein bottle amplitudes.)

The construction of correlators consists in a beautiful combination
of the construction of \cite{fffs2,fffs3} with structures familiar from
lattice TFTs in {\em two\/} dimensions. As is already apparent from our
prescription for the torus and annulus amplitudes, what is needed is
a Wilson graph in the connecting three-manifold $M_X$, which
besides the Wilson lines that correspond to field insertions and to the
presence of a boundary contains in addition a suitable network of $A$-lines.
The general prescription is the following:\\[.4em]
(0) To start, we fix once and for all an orientation of the world sheet $X$.
    The complex double $\hat X$ of $X$ can then be decomposed as $\hat X\eq
    X^+_{} {\sqcup} \partial X {\sqcup} X^-_{}$, with the involution acting as
    $\sigma(X^+_{})\eq X^-_{}$ and $\sigma(\partial X)\eq\partial X$.\\[.2em] 
(1) Each bulk insertion point $p_\ell$ on $X$ has two preimages on $\hat X$,
    $\hat p_\ell^+$ on $X^+$ and $\hat p_\ell^-$ on $X^-$, which are joined 
    by an interval in the connecting manifold $M_X$. We put a Wilson line
    along each such interval. The image of $X$ in $M_X$ intersects the
    interval in a unique point, which we identify with $p_\ell\iN X$. \\[.2em]
(2) Each component of the boundary of $X$ gives rise to a line of fixed points
    of $M_X$ under the $\zet_2$ action. We place a circular Wilson line along
    each such line of fixed points. \\[.2em]
(3) Boundary insertion points $q_\ell$ on $X$ have a unique preimage
    $\hat q_\ell$ on $\hat X$. We join $\hat q_\ell$ by a short Wilson line to
    the image of $q_\ell$ in $M_X$. This results in a trivalent vertex on the
    circular Wilson line for the relevant boundary component. \\[.2em]
(4) We pick a triangulation of the image of $X$ in $M_X$. Without loss of
    generality, we do this in the following manner. First, only trivalent
    vertices may appear. (But we allow for  arbitrary polygonal faces rather
    than just triangles, which is completely  equivalent \cite{chfs} to the
    case with only triangular faces; for brevity we still use the term
    `triangulation'). Moreover, each segment of the boundary must contain the
    end point of an edge of the triangulation, and each of the bulk points 
    $p_\ell$ must lie in the interior of a separate face of the triangulation,
    while each boundary point $q_\ell$ must lie in the interior of a separate
    edge. Finally, for each bulk insertion we join the perpendicular bulk
    Wilson line in $p_\ell$ by an additional line to an interior point of an
    arbitrary edge of the face to which $p_\ell$ belongs.
    (More precisely, in terms of ribbons, we first bend the `horizontal' ribbon
    straight up, at a right angle to $X$, so that it points towards $p_\ell^-$
    and is parallel to the vertical bulk ribbon, and then join it to the bulk
    ribbon.)
\\[.4em]
This completes the geometric part of our prescription. It is illustrated,
in the case of a disk with three bulk and three boundary insertions, 
in the following figure.
  \be  \begin{picture}(177,151)(0,0)
  \put(0,0)   {\begin{picture}(0,0)(0,0)
              \scalebox{.38}{\includegraphics{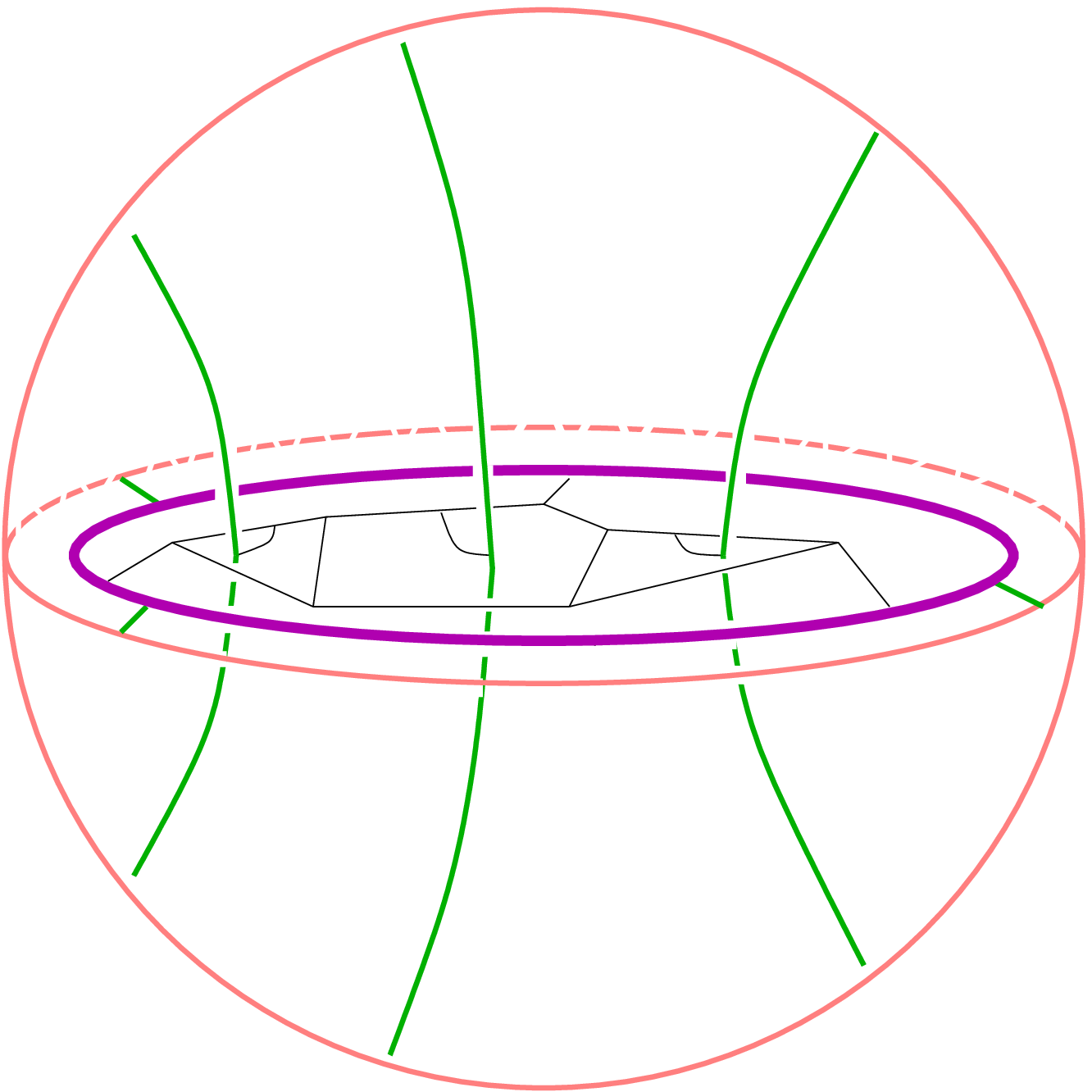}} \end{picture}}
  \end{picture} \labl{disk0}
The remaining task is now to
decorate the Wilson lines with labels specifying the corresponding
object of the category and choose couplings (morphisms) for the trivalent
vertices. Our prescription is inspired by the situation in \twodim\ lattice 
TFTs and contains that case as a special example.
\\[.4em]
(5) The short Wilson lines that connect the bulk insertion points $p_\ell$
    to the triangulation are labelled by the algebra $A$, with the $A$-line 
    pointing towards $p_\ell$. \\[.2em]
(6) The three lines that join at any vertex of the triangulation
    are interpreted as outgoing and are labelled with the algebra object $A$.
    The morphisms at the vertices are constructed from the coproduct in $A$:
    At every trivalent vertex at which these $A$-lines join we put the
    morphism $\Delta\cir\Phi_1^{-1}$, with $\Phi_1$ given by \erf{Phi12}.
    (The condition that $A$ is symmetric ensures that the resulting correlators
    do not depend on the choice of which of the three lines carries the
    $\Phi_1^{-1}$.) In the lattice case, this prescription corresponds to the
    assignment of structure constants with only lower indices.\\[.2em]
(7) In the case of lattice TFT one also needs a metric. In the general case
    this corresponds to reversing the direction of the $A$-line. Accordingly
    we place the morphism \erf{Phi12} in $\Hom(A,A^\vee)$ in the middle of
    each edge of the triangulation. \\[.2em]
(8) To characterize the bulk field inserted at $p_\ell$, we need two
    irreducible objects $j_\ell^\pm$; they label the Wilson lines that start
    from the points $\hat p_\ell^\pm$ on $\hat X$. To account for the coupling
    to the short $A$-lines, we need as a third datum for a bulk field a
    morphism in $\Hom(A\oti j_\ell^+,j_\ell^{-\vee})$. \\[.2em]
It turns out that some of the latter morphisms completely decouple, i.e.\
that every correlator containing a bulk field labelled by such a coupling 
vanishes. Physical bulk fields therefore correspond to a subspace of 
the couplings $\Hom(A\oti j_\ell^+,j_\ell^{-\vee})$; its dimension matches the 
value $Z_{j_\ell^+j_\ell^{-\vee}}$ of the torus partition function.
 \\[.2em]
(9) The segments of the circular Wilson line correspond to boundary conditions;
    they are to be labelled by a module of $A$. Wilson lines of the
    triangulation that end on such a boundary result in a trivalent vertex,
    which requires a morphism in $\Hom(A\oti\M,\M)$. For this morphism we choose
    the \rep\ morphism $\rho_M$. \\[.2em]
(10) The last ingredient in our construction are the boundary fields. They have
    a single chiral label $k_\ell$, which we use to label the short Wilson line
    from $\hat q_\ell$ to $q_\ell$. The trivalent vertex that is formed by this
    Wilson line and the two adjacent boundary conditions $M,N$ requires the
    choice of a coupling in $\Hom(M\oti k_\ell,N)$. \\[.2em]
Again, only a subspace of these couplings is relevant. In this case the 
subspace can easily be characterized: $M\oti k_\ell$ is a left $A$-module, and
only the subspace $\Hom_A(M\oti k_\ell,N)$, of dimension ${\rm A}_{k_\ell M }
^{\ \ \ N}$, of $A$-morphisms gives rise to physical boundary fields that 
change the boundary condition from $M$ to $N$.
\\[.4em] 
In the example of the disk, the picture then looks as follows
(after rearranging, in one of several possible ways, the vertices that
result from the presence of $\Phi_1$ and its inverse, so as to eliminate
all occurrences of unit and co-unit):
  \be  \begin{picture}(177,157)(0,0)
  \put(0,19)  {$$}
  \put(0,0)   {\begin{picture}(0,0)(0,0)
              \scalebox{.38}{\includegraphics{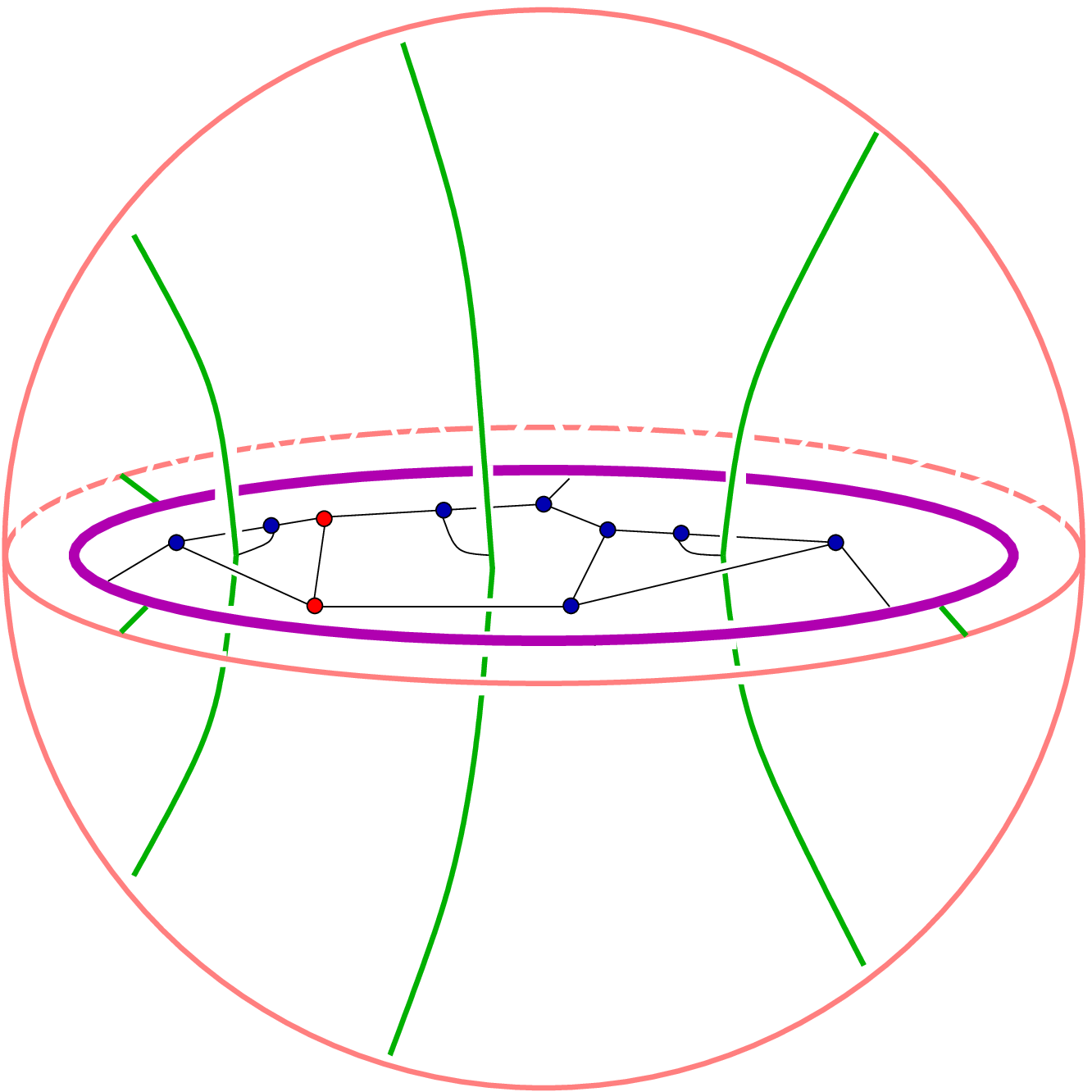}} \end{picture}}
  \put(-3.2,70) {\scriptsize$M_3$}
  \put(11.5,55) {\scriptsize$k_1$}
  \put(12,85.6) {\scriptsize$k_3$}
  \put(20.7,29) {\scriptsize$j_1^+$}
  \put(22.6,112){\scriptsize$j_1^{-\vee}$}
  \put(55.4,8)  {\scriptsize$j_2^+$}
  \put(58.4,134){\scriptsize$j_2^{-\vee}$}
  \put(77,53.7) {\scriptsize$M_1$}
  \put(111,82)  {\scriptsize$M_2$}
  \put(112.4,28){\scriptsize$j_3^+$}
  \put(111,111) {\scriptsize$j_3^{-\vee}$}
  \put(127,54){\scriptsize$k_2$}
  \end{picture} \labl{disk}

\vskip.4em

Topological invariance of a \twodim\ lattice model means that all \corfu s are
independent of the choice of triangulation. The same type of arguments as in
the topological case can be used to deduce independence from the triangulation
also in the CFT case. Concretely, triangulation independence can be reduced (in
a dual formulation) to invariance under two local moves, the so-called {\em 
fusion\/} and {\em bubble\/} moves \cite{fuhk,chfs}. These moves look like
  \be  \begin{picture}(260,69)(0,0)
  \put(0,0)   {\begin{picture}(0,0)(0,0)
              \scalebox{.38}{\includegraphics{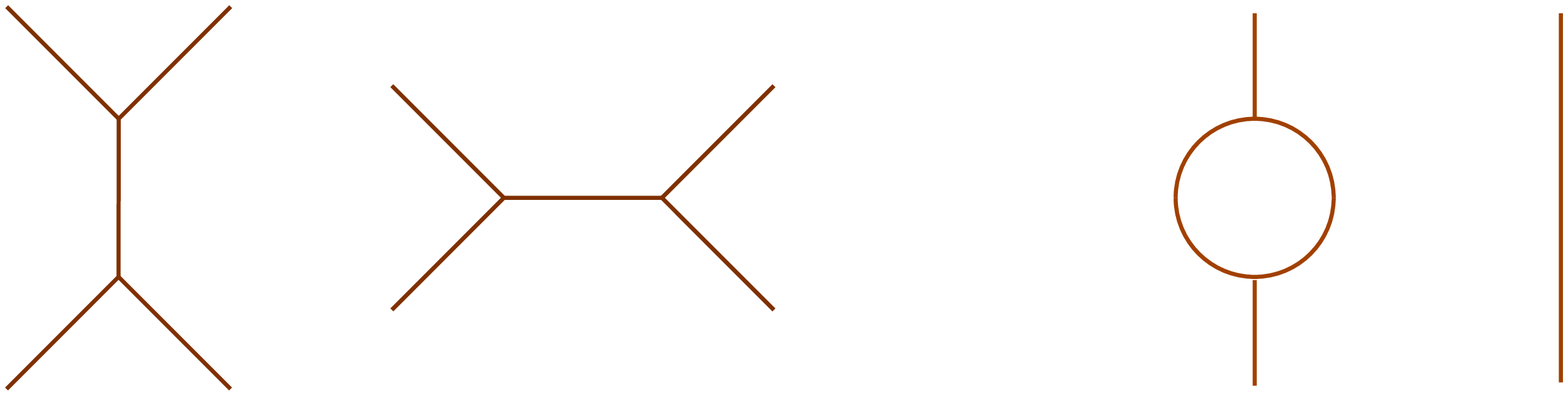}} \end{picture}}
  \put(46,31) {$\leftrightarrow$}
  \put(161,31){and}
  \put(244,31){$\leftrightarrow$}
  \end{picture} \labl{bubble}
Our construction amounts to regard the lines in this picture as morphisms and 
decorate all of them with the label $A$, and to interpret the trivalent vertices 
as products and coproducts, respectively. The fusion move is, in CFT terms, 
just crossing symmetry between the s- and t-channel, and hence invariance under
this move is guaranteed by the associativity and the Frobenius property 
\erf{frobenius} of $A$.
Similarly, the second the equality \erf{bubble} is nothing but the second 
part of the formula \erf5, i.e.\ is guaranteed by the fact that $A$ is special.
Together with the naturality properties of 3-d topological field theory,
the independence from the triangulation implies that the correlators are
invariant under the relative modular group of \cite{bisa2}. Furthermore, the 
correct factorization rule for boundary fields follows directly from dominance
properties of the category $\calc$. Bulk factorisation requires in addition a 
surgery move on the connecting three-manifold. 

\section{Morita equivalence and T-duality}

We have shown how to obtain a full conformal field theory from a symmetric
special Frobenius algebra $A$ in the modular tensor category of the underlying
chiral CFT. The relation between such algebras and full \cfts\ is, however,
not one-to-one. Rather, different algebras can describe one and the same 
conformal field theory. In other words, the algebra $A$ itself should not 
be thought of as an observable quantity.

This is already familiar from the degenerate realization of our construction 
that is provided by topological lattice theories. In that case, symmetric 
Frobenius algebras over the complex numbers with the same number of simple 
ideals (and thus isomorphic centers) give identical theories \cite{fuhk}. 
In particular, the algebra can therefore be chosen to be commutative.

The general case is treated as follows. We call two 
symmetric special Frobenius algebras $A$ and $B$ {\em Morita equivalent\/} iff
there exist two bimodules $_A M_B$ and $_B M'_A$
(the first a left module of $A$ and a right module of $B$, the second a
left module of $B$ and a right module of $A$) such that
  \be (_A M_B) \otimes_B^{} (_B M'_A) = A \qquad\mbox{and}\qquad
  (_B M'_A) \otimes_{\!A}^{} (_A M_B) = B \,.  \labl{mor}
Using these relations, one can replace a triangulation of the world sheet
$X$ that is labelled by $A$-lines with the dual triangulation labelled by 
$B$-lines. Moreover, by standard arguments \cite{Vari} it follows that Morita
equivalent algebras have the same representation theory, so that $A$-modules
and $B$-modules, and hence boundary conditions, are in one-to-one
correspondence. Since the correlation functions do not depend on the
triangulation, this implies that our prescription yields, upon a relabelling 
of fields and boundary conditions, the same correlators when performed 
with either of the two algebras $A$ and $B$.

These observations also allow for a general derivation
of T-duality in rational CFT. Suppose we are given a `symmetry' $\omega$ of 
the chiral data. Technically, this is a group $G$ together with a functor from
$\calc[G]$ -- a category whose objects are pairs of objects of $\calc$ and 
an action of $G$ by automorphisms of the objects -- to $\calc$; it includes
in particular a permutation $i\,{\mapsto}\,\omega i$ of the primary fields such 
that $\omega\one\eq\one$, $T_{\omega i}\eq T_i$ and $S_{\omega i\,\omega j}
\eq S_{ij}$. It is straightforward to check that along with $Z_{ij}$ also 
$Z_{i\,\omega j}$ is a modular invariant partition function. (A familiar 
realization is the standard T-duality in the CFT of a free boson. The 
relevant symmetry is the one that reverses the sign of the U$(1)$ charges.)
The idea is to associate to the symmetric special
Frobenius algebra $A$ that gives the partition function $Z_{ij}$ another
algebra $A^\omega$ that yields the partition function $Z_{i\,\omega j}$.
The construction is based on Morita equivalence in the orbifold theory
$\calc^\omega$ that is obtained by modding out the symmetry $\omega$ on
$\calc$. One can show that with respect to this orbifold theory, the algebras
$A$ and $A^\omega$ give, upon appropriate identification of the fields and 
of the \bc s, isomorphic correlation functions on all surfaces.

\section{Conclusions}

We have shown how to obtain full rational conformal field
theories based on the chiral data that are encoded in a modular tensor
category $\calc$ from (Morita equivalence classes of) symmetric
special Frobenius algebras in $\calc$. We gave a general prescription
for the construction of \corfu s on orientable surfaces, including
surfaces with boundary. We will demonstrate elsewhere that the so 
obtained correlators give rise to a fully consistent CFT.

Our results open up several lines of research. First, the generalization of
these results to unorientable surfaces, a necessary step to a deeper
understanding of type I string compactifications, will require
more structure that has to account, in particular, for the different
possible choices of orientifold projections.

Another important goal is classification. In the framework we propose, the 
problem of classifying rational \cfts\ can be divided in three
independent tasks, each of which deals with a clearly posed (albeit very
difficult) problem. The first is to classify modular tensor categories.
Second, given a modular tensor category \calc, two distinct problems
must be attacked. On one hand we must determine the Morita equivalence 
classes of symmetric special Frobenius algebras in \calc. Note that this 
issue can be studied entirely at a `topological' level, i.e.\ no analytic
properties of chiral blocks are involved. On the other hand, one also 
needs to get a handle on the control data that allow to completely 
reconstruct a chiral CFT from its modular tensor category. (The 
correspondence between chiral CFT and modular tensor categories is not
one-to-one. For example, all WZW theories based on $so(2n{+}1)$ at level 1 
with values of $n$ that differ by a multiple of 24 have equivalent tensor 
categories.) This part of the problem is a purely chiral issue. It
is closely related to the classification of chiral algebras, which at
present remains the least accessible part of the classification programme.

In order that this programme is complete, we also still must establish
that every full rational CFT can be obtained from some symmetric
special Frobenius algebra $A$. In all examples known to us -- in particular
for all pure extensions as well as for all simple current modular invariants
-- this is indeed the case, but at present we cannot give a general proof.
What is desirable is a general prescription for reconstructing the
algebra object $A$ from some collection of \corfu s of the full CFT.
It would also be interesting to see whether this reconstruction is related to 
other algebraic structures, such as the double triangle algebras \cite{ocne7} 
which in the work of \cite{pezu6,pezu8} are regarded as a structure underlying 
every rational CFT. This would also allow for a comparison between the
expressions for correlators obtained there and those which follow from our
prescription.

We are confident that the framework we propose is flexible enough to allow
for an extension to non-rational conformal field theories. In fact, rationality
implies that the category $\calc$ is semi-simple. Generalizations of
TFT for non semi-simple tensor categories have been studied in the literature, 
compare the recent book \cite{KElu}. 
It can be expected that the concepts which underlie these generalizations will
play a role in the analysis of non-rational theories. Finally, we point out
that our results suggest an intimate relation between the problem of deforming
algebras and the problem of deforming conformal field theories.

\vspace{3em}\noindent
{\bf Acknowledgments}:\\[1mm] We are grateful to Giovanni Felder and Albert Schwarz
for helpful discussions and comments, and to the Erwin Schr\"odinger Institute 
and to Karlstads universitetet for hospitality.
IR is supported by the EU grant HPMF-CT-2000-00747. 

\newpage\noindent
{\bf Note added in proof}:\\[3mm]
In \cite{ostr}, which appeared after publication of the present paper,
a reconstruction theorem for module categories is proven. 
When combined with the general expectation that
boundary conditions of a rational conformal field theory
lead to a module category over the category $\calc$
that describes the Moore\hy Seiberg data of the chiral CFT, the results of
\cite{ostr} provide an additional argument for the existence of an
algebra object $A$, and thereby for the generality of the approach proposed in 
the present paper.
\vspace{3em}


\newcommand\wb{\,\linebreak[0]} \def\wB {$\,$\wb}
 \newcommand\Bi[1]    {\bibitem{#1}}
 \newcommand\J[5]     {{\sl #5\/}, {#1} {#2} ({#3}) {#4} }
 \newcommand\K[6]     {{\sl #6\/}, {#1} {#2} ({#3}) {#4}}
 \newcommand\Prep[2]  {{\sl #2\/}, pre\-print {#1}}
 \newcommand\BOOK[4]  {{\sl #1\/} ({#2}, {#3} {#4})}
 \newcommand\inBO[7]  {{\sl #7\/}, in:\ {\sl #1}, {#2}\ ({#3}, {#4} {#5}), p.\ {#6}}
 \def\jf    {J.\ Fuchs}
 \def\dim   {dimension}  
 \def\comp  {Com\-mun.\wb Math.\wb Phys.}
 \def\cpma  {Com\-pos.\wb Math.}
 \def\fiic  {Fields\wB Institute\wB Commun.}
 \def\foph  {Fortschritte\wB d.\wb Phys.}
 \def\ijmp  {Int.\wb J.\wb Mod.\wb Phys.\ A}
 \def\jktr  {J.\wB Knot\wB Theory\wB and\wB its\wB Ramif.}
 \def\jomp  {J.\wb Math.\wb Phys.}
 \def\nupb  {Nucl.\wb Phys.\ B}
 \newcommand\phgt[2] {\inBO{Physics, Geometry, and Topology}
            {H.C.\ Lee, ed.} \PL\NY{1990} {{#1}}{{#2}} }
 \def\phep  {Proc.\wb HEP$\!$}
 \def\phlb  {Phys.\wb Lett.\ B}
 \def\phrl  {Phys.\wb Rev.\wb Lett.}
 \def\slnm  {Sprin\-ger\wB Lecture\wB Notes\wB in\wB Ma\-the\-matics }
 \def\AMS    {{American Mathematical Society}}
 \def\PL     {{Plenum Press}}
 \def\SV     {{Sprin\-ger Ver\-lag}}
 \def\WI     {{Wiley Interscience}}
 \def\Be     {{Berlin}}
 \def\PR     {{Providence}}
 \def\NY     {{New York}}


\end{document}